 \documentclass[final,5p,times,twocolumn]{elsarticle}

\usepackage{subfigure}
\usepackage{amssymb}
\usepackage{lipsum}
\usepackage{amsthm}
\usepackage{graphicx}
\usepackage{amsmath}
\usepackage[colorlinks=true, linkcolor=blue, citecolor=blue, urlcolor=blue]{hyperref}

\journal{ }

\begin{document}

\begin{frontmatter}



    \title{Einstein-Horndeski gravity and the ultra slowly evaporating black hole}


\author[first]{Xiao Liang}
\ead{xliang@nuaa.edu.cn}
\author[first]{Yu-Sen An}
\ead{anyusen@nuaa.edu.cn}
\author[first]{Chen-Hao Wu}
\ead{chenhao\_wu@nuaa.edu.cn}
\author[first,second]{Ya-Peng Hu\fnref{label1}}
\fntext[label1]{Corresponding author}
\ead{huyp@nuaa.edu.cn}
\affiliation[first]{organization={College of Physics, Nanjing University of Aeronautics and Astronautics, Nanjing, 210016, China}}
\affiliation[second]{organization={MIIT Key Laboratory of Aerospace Information Materials and Physics,  Nanjing University of Aeronautics and Astronautics, Nanjing, 210016, China}}

\begin{abstract}
In this work, we study the evaporation behaviors of asymptotically flat charged black holes in the Einstein-Horndeski gravity theory. Based on the thermodynamics of the Horndeski black hole, we present a physical understanding of the scalar charge of the Horndeski black hole and also clarify
its connection to the Einstein vector theory. As the presence of non-minimal coupling, the evaporating behaviors of the Horndeski black hole are vastly different from the Reissner-Nordstrom (RN) black hole case. Due to the different spacetime and electric field structures, the evaporation rate of the Horndeski black hole will slow down at the late stage of evaporation and thus gain a lifetime much longer than the RN black hole. These results illuminate the effect of non-minimally coupled matters on the black hole evaporation and provide clues to search for these matter fields in future observations.
\end{abstract}



\begin{keyword}
black hole \sep evaporation \sep scalar-tensor theory 



\end{keyword}

\end{frontmatter}




\section{Introduction}
Through indirect and direct observations by LIGO and the Event Horizon Telescope \cite{LIGOScientific:2016aoc, LIGOScientific:2017ycc, EventHorizonTelescope:2019dse, EventHorizonTelescope:2022wkp}, black holes, the most extraordinary objects in the universe, have received increasing attention. One of the most interesting things about black holes is that they behave like a thermodynamical system. In Ref.\citep{Hawking:1974rv, Hawking:1975vcx}, by using quantum field theory in curved space-time, Hawking showed that quantum effects allow black holes to radiate like black bodies which are now called Hawking radiation. From the Hawking radiation spectrum, one can calculate the power of the particle emission, investigate the evaporation behavior, and estimate the lifetime of black holes. After the discovery of Hawking radiation, Page gave numerical results for the lifetimes of uncharged and non-rotating black holes in asymptotically flat spacetime \citep{Page:1976df}. He found that there is a simple relation between a black hole's lifetime and its initial mass $t \sim M_0^3$ \citep{Page:1976df}, and then extended these discussions to rotating black hole case \citep{Page:1976ki}. In Ref.\citep{Hiscock:1990ex}, by combining the Hawking radiation and Schwinger effect in quantum electrodynamics, Hiscock and Weems generalized the discussions of black hole evaporation to the charged black hole case. 
They found that the presence of electric charge may greatly extend the lifetime of a black hole \citep{Hiscock:1990ex}. The study of the evaporation behavior of black holes has been a fruitful area up until now, it has been extended to many different spacetime backgrounds, such as de-Sitter spacetimes \citep{Gregory:2021ozs} and anti-de-Sitter spacetimes \citep{Page:2015rxa}. Due to the diversity of gravitational theories, there are also many discussions on black hole evaporation in modified gravity theories \citep{Wu:2021zyl, Hou:2020yni, Xu:2020xsl,Xu:2019krv,An:2024fzf}. Moreover, black hole evaporation may also be of help in the search for dark matter and dark energy \citep{Liang:2023jrj}.

Einstein's general relativity where all the matter fields are minimally coupled to gravity has passed various direct or indirect tests. However, many extensions of general relativity
have also been taken into account in order to explain the accelerated expansion of our universe \citep{SupernovaSearchTeam:1998fmf, SupernovaCosmologyProject:1998vns}. Particularly, a lot of work has been devoted to finding the most general non-minimal coupled scalar-tensor action whose
equations of motion have at most second-order derivatives.
In the 1970s,  general classes of theories of this kind were constructed by Horndeski, which is now called Einstein-Horndeski gravity \cite{Horndeski:1974wa}. Although Horndeski gravity is a scalar-tensor theory, by gauging the shift symmetry of the scalar field, it is also equivalent to the Einstein-vector theory which is shown in Ref.\citep{Geng:2015kvs}. Horndeski gravity has the coupling between the derivative of the scalar field and Einstein tensor. This feature makes Horndeski gravity theory useful in inflationary cosmology \cite{Sushkov:2009hk, Germani:2010hd}
and late-time cosmology \cite{Saridakis:2010mf}. Besides investigations in the cosmological background, Horndeski gravity has also received much attention in static and spherically symmetric black hole systems. Various interesting topics have been discussed in Horndeski black hole background. To name a few, Ref.\citep{Hu:2018qsy, Miao:2016aol, Feng:2015wvb} discussed the Horndeski black hole thermodynamics, Ref.\citep{Feng:2015oea, Kuang:2016edj, Feng:2018sqm} discussed the holographic implication of Horndeski black hole, Ref.\citep{Jha:2022tdl} focused on superradiation of Horndeski black hole and \citep{Gao:2023mjb} discussed the black hole shadow in Horndeski background.  Despite the above progress,  black hole evaporation behavior which is an important part of black hole thermodynamics has not been investigated in Horndeski gravity background. 

Thus in this paper, we will focus on the evaporation behavior of asymptotically flat charged black holes in Einstein-Horndeski theory.  However, in astrophysics, the role of electric charge is usually considered insignificant due to the neutralization of the galactic medium. The possible presence of vector type dark matter may still result in a charged black hole. Thus we still choose the charged black hole as our research object. 
The structure of this paper is as follows: In Sec. 2, we give a review of Einstein-Horndeski theory and the associated asymptotically flat black hole solution. The novel properties of this black hole solution are also briefly discussed. In Sec.3, we briefly introduce the thermodynamic relations of this Einstein-Horndeski black hole before investigating its evaporation behavior. In Sec.4, after giving the physical meaning of scalar charge in the Horndeski black hole, we numerically calculate the evaporation behavior of this black hole. Finally, we conclude our paper in Sec.5. Throughout this paper, we use the natural units with $4\pi\epsilon_0=\hbar=k_B=G =c=1$.

\section{Einstein-Horndeski theory and charged black hole solution.}
In this section, we first review the properties of asymptotically flat charged black hole solution in Horndeski gravity. The action of Horndeski gravity with a $U(1)$ gauge field is given by \citep{Feng:2015wvb}
\begin{equation}
    S=\int d^4x \sqrt{-g}\bigg[(R-2\Lambda-\frac{1}{4}F^2)-\frac{1}{2}(\alpha g_{\mu\nu}-\eta G_{\mu\nu})\nabla^\mu\phi\nabla^\nu\phi\bigg], 
\end{equation}
where $\alpha$ is a dimensionless coupling constant, the $\eta$ represents the coupling constant between Einstein tensor and scalar field whose dimension is $[L]^{2}$.  $F=dA$ is the U(1) gauge field strength and the field $\phi$ represents a non-minimally coupled real scalar field. The scalar field is axionic and only appears through its derivative in the action, so this theory is invariant under a constant shift of $\phi$. From the above action, one can get the equation of motion as follows
\begin{equation}
\begin{aligned}
    &G_{\mu\nu}+\Lambda g_{\mu\nu}=\frac{1}{2}(\alpha T_{\mu\nu}+\eta\Xi_{\mu\nu}+\frac{1}{2}F^2_{\mu\nu}-\frac{1}{8}F^2 g_{\mu \nu}),\\
    &\nabla_{\mu}[(\alpha g^{\mu\nu}-\eta G^{\mu\nu})\nabla_{\nu}\phi]=0,\\
    &\nabla_\nu F^{\nu \mu}=0,\label{eom1}
\end{aligned}
\end{equation}
where $T_{\mu\nu}$ and $\Xi_{\mu\nu}$ can be written as
\begin{equation}
    \begin{aligned}
    T_{\mu\nu} &= \nabla_{\mu}\phi\nabla_{\nu}\phi-\frac{1}{2}g_{\mu\nu}\nabla_{\rho}\phi\nabla^{\rho}\phi,\\
    \Xi_{\mu\nu} &\equiv \frac{1}{2}\nabla_{\mu}\phi\nabla_{\nu}\phi R-2\nabla_{\rho}\phi\nabla_{(\mu}\phi R_{\nu)}^{\rho}-\nabla^{\rho}\phi\nabla^{\lambda}\phi R_{\mu\rho\nu\lambda} & \\  & -(\nabla_{\mu}\nabla^{\rho}\phi)(\nabla_{\nu}\nabla_{\rho}\phi)+(\nabla_{\mu}\nabla_{\nu}\phi)\square\phi+\frac{1}{2}G_{\mu\nu}(\nabla\phi)^2 & \\  & -g_{\mu\nu}\bigg[-\frac{1}{2}(\nabla^{\rho}\nabla^{\lambda}\phi)(\nabla_{\rho}\nabla_{\lambda}\phi)    +\frac{1}{2}(\Box\phi)^2-\nabla_{\rho}\phi\nabla_{\lambda}\phi R^{\rho\lambda}\bigg]. & 
    \end{aligned}
\end{equation}
Interestingly, from the equation of scalar field, a conserved current can be defined as 
\begin{equation}
    J^{\mu}=(\alpha g^{\mu\nu}-\eta G^{\mu\nu})\nabla_{\nu}\phi, \label{shl}
\end{equation}
which can be used to prove the no-hair theorem as shown in Ref.\citep{Hui:2012qt}.
In this work, we only consider the static and spherically symmetric black hole solution. So we can write the most general ansatz as
\begin{equation}
\begin{aligned}
    ds^{2}&=-h(r)dt^{2}+\frac{dr^{2}}{f(r)}+r^{2}d\Omega_{2}^{2},\\
    \phi&=\phi(r),\hspace{1cm} A=\Phi(r)dt.
    \end{aligned}\label{anz}
\end{equation}
in which the $\Phi(r)$ is the U(1) gauge field potential. By solving the Maxwell equation $\nabla_\nu F^{\nu \mu}=0$, the field potential $\Phi$ can be expressed as
\begin{equation}\Phi^\prime=\frac{4Q}{r^2}\sqrt{\frac{h}{f}},
\end{equation}
where $Q$ is an integration constant. By plugging the ansatz (\ref{anz}) into the above equations (\ref{eom1}), there are three more independent equations which read
\begin{equation}
    \begin{aligned}
    \left(\sqrt{\frac{f}{h}}(2\eta rfh^{\prime}+2\eta(f-1)h-2\alpha r^2h)\phi^{\prime}\right)^{\prime}  &=0,\\
    4(2rf^{\prime}+2(f-1)+2\Lambda r^2)  +\frac{32Q^2}{r^2}+2\alpha r^2f\phi^{\prime2}\\  +2\eta(4rf\phi^{\prime\prime}+(3rf^{\prime}+(f+1))\phi^{\prime})f\phi^{\prime} &=0,\\
     4(2rfh^{\prime}+2h(f-1)+2\Lambda r^2h)+\frac{32Q^2}{r^2}h-2\alpha r^2fh\phi^{\prime2}  \\   +2\eta(3rfh^{\prime}+(3f-1)h)f\phi^{\prime2} &=0. 
     \end{aligned}\label{eom}
\end{equation}
The most general analytical solutions of these equations would be difficult to obtain explicitly.  A special class of solutions was obtained in \citep{Cisterna:2014nua} by taking the specific condition 
\begin{equation}
    \eta(2rfh^\prime+2(f-1)h)-2\alpha r^2h=0.\label{fe}
\end{equation}
It should be pointed out that this equation is equivalent to $\alpha g^{rr}-\eta G^{rr}=0$, which is obviously related to Eq.(\ref{shl}). The underlying physical meaning will be discussed later. 

For the case $\alpha \neq 0$ and $\eta \neq 0$, by analyzing the Riemann tensor near the spacetime infinity,
the AdS radius can be defined as
\begin{equation}
    l^{-2}:=\frac{\alpha}{3\eta},
\end{equation}
where $\alpha/\eta$ is positive by the requirement of reality of the solution \citep{Cisterna:2014nua}. Thus for non-zero $\alpha$, the spherical black hole spacetime can only be the asymptotically AdS black hole. 
However in this paper, based on the observational consideration, we are more interested in asymptotically flat black hole solution. In order to get asymptotically flat black hole, we should set $\alpha=0$ which corresponds to the divergent AdS radius. So we will focus on the case where $\alpha=0$ and $\Lambda=0$ in the following. In this scenario, by directly solving the coupled differential equations Eq.(\ref{eom}) and (\ref{fe}), an asymptotically flat black hole solution can be obtained where \citep{Cisterna:2014nua, Feng:2015wvb}
\begin{equation}
\begin{aligned}
  &h(r)=1-\frac{2M}{r}+\frac{4Q^2}{r^2}-\frac{4Q^4}{3r^4},\\
    &f(r)=\frac{r^4}{(r^2-2Q^2)^2}h(r),\\
   &\Phi=\Phi_{e0}-\frac{4Q}{r}+\frac{8Q^3}{3r^3},\\& \phi'=\sqrt{-\frac{8Q^{2}}{\eta r^{2}f}}.
     \end{aligned} \label{solution}
\end{equation} 
There are three integration parameters $\Phi_{e0}$, $M$ and $Q$. We will choose the infinity as the reference point of electric potential and thus set $\Phi_{e0}=0$. The $M$ is the mass of the black hole and $Q$ stands for its gauge field charge. From here we can see that in the asymptotically flat case, the existence of the scalar field depends on the charge $Q$ and the coupling constant $\eta$. 

There are two singularities in (\ref{solution}), one is at $r=0$ , and the other is $ r= r_*\equiv \sqrt{2}Q$. Both of them lead to the divergence of Kretschmann scalar $K=R_{\mu\nu\rho\sigma}R^{\mu\nu\rho\sigma}$. By weak cosmic censorship conjecture which forbids the naked singularity,  the event horizon radius must be larger than the singularity $r_*$. This condition implies that the mass and charge of this Horndeski black hole must follow the relation
\begin{equation}
    \frac{Q}{M}<\frac{3\sqrt{2}}{8}.\label{cons}
\end{equation} 
Moreover, from the metric, it is clear to see that this metric will reduce to the Schwarzschild black hole when $Q =0$, but it can not reduce to the RN black hole in the limit $\eta \to 0$ as the solution only exists for parameter $\eta\neq 0$.

\begin{figure}[h]
    \centering
    \includegraphics[width=0.95\linewidth]{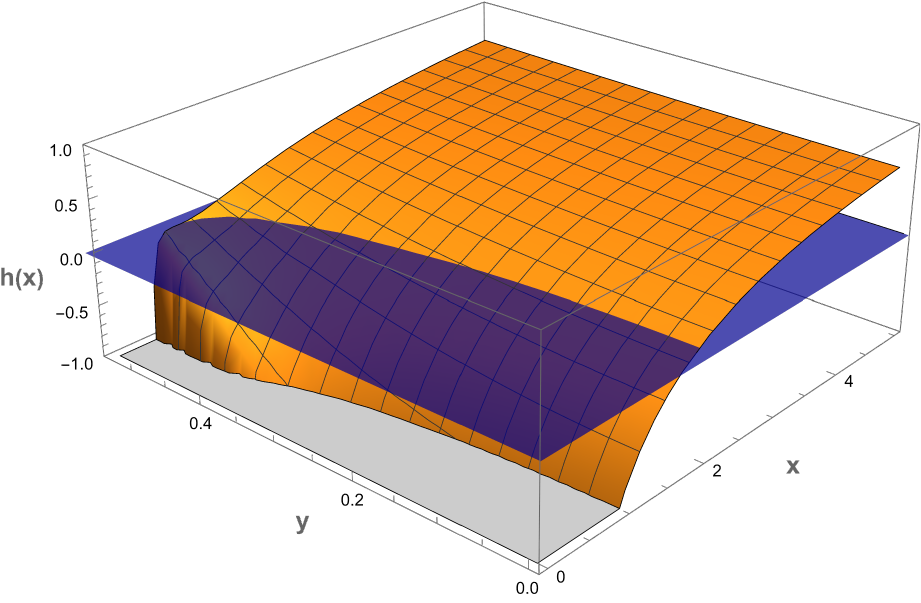}
    \caption{The diagram of the metric function $f(r)$ varying with respect to $Q$ and $r$. In this plot, $Q/M$ is   constrained by cosmic censorship conjecture ranging from $0$ to $\frac{3\sqrt{2}}{8}$.This figure shows that the black hole only has one horizon.}
    \label{FR}
\end{figure}

One can obtain the event horizon by calculating the roots of the metric function $h(r)$
\begin{equation}\label{hori}
   h(r_{h})=1-\frac{2M}{r_h}+\frac{4Q^2}{r_h^2}-\frac{4Q^4}{3r_h^4}=0 .
\end{equation}
The metric function $h(r)$ can also be expressed in terms of dimensionless parameters $x=r/M$ and $y=Q/M$ as 
\begin{equation}
  h(x)=1-\frac{2}{x}+\frac{4y^2}{x^2}-\frac{4y^4}{3x^4}.
\end{equation}
The charge parameter $y$ of this Horndeski black hole always satisfies the constraint Eq.(\ref{cons}). Under this constraint, we plot the $h(r)$ in Fig.\ref{FR} to show the black hole metric function for different values of charge parameter $y$. It is easy to find that the metric function has only one line intersecting the $z=0$ plane which means that there is only one positive real root of Eq.(\ref{hori}). This means the charged Horndeski black hole we considered has only one event horizon. It should be stressed that this behavior with only one horizon is different from cases in Einstein gravity where the charged black holes and rotating black holes both have two horizons \citep{Brown:2015bva}. Interestingly, the rotating black hole in Horndeski gravity can also have both inner and outer horizons as shown in Ref.\citep{Santos:2024zoh}. 


\section{Thermodynamics of charged black holes in Einstein-Horndeski gravity}
Before discussing the evaporation of Horndeski black hole, we must know the thermodynamics of the system first. So we will briefly introduce how the thermodynamics of charged Einstein-Horndeski black holes is obtained in this section following \citep{Feng:2015wvb}.

For the asymptotical flat charged black hole in Einstein-Horndeski gravity, the Hawking temperature can be calculated by using the Euclidean method
 \begin{equation}
    \left.T=\frac{\sqrt{f'(r) h'(r)}}{4\pi}\right|_{r=r_h}=\frac{r_h^2-2Q^2}{4\pi r_h^3}.\label{temp}
 \end{equation}
Note that the condition $r_{h}>\sqrt{2}Q$ also guarantees the temperature to be positive. 

The first law of thermodynamics in Horndeski case follows from the Wald formalism which is a result of the following simple relation \citep{Feng:2015wvb}
\begin{equation}
    \int_{c} \omega(\tilde{\phi},\delta \tilde{\phi},\mathcal{L}_{\xi}\tilde{\phi})=\delta \mathcal{H}_{\infty}-\delta \mathcal{H}_{+},
\end{equation}
where $\tilde{\phi}$ denotes all kinds of fields and $\omega(\tilde{\phi},\delta \tilde{\phi},\mathcal{L}_{\xi}\tilde{\phi})$ is called symplectic current which is a linear function in terms of $\mathcal{L}_{\xi}\tilde{\phi}$. For static spherically symmetric spacetime, there is a killing vector that is null on the horizon. For $\xi$ to be killing vector, $\mathcal{L}_{\xi}\tilde{\phi}=0$, which leads to $\omega(\tilde{\phi},\delta\tilde{\phi},\mathcal{L}_{\xi}\tilde{\phi})=0$, and thus the relation $\delta \mathcal{H}_{\infty}=\delta \mathcal{H}_{+}$ holds. 
As computed in Ref.\citep{Feng:2015wvb}, the $\delta \mathcal{H}_{\infty}$ is 
\begin{equation}
    \delta \mathcal{H}_{\infty}=\delta M-\Phi_{e}\delta Q,
\end{equation}
while the term at horizon $\delta H_{+}$ reads 
\begin{equation}
    \delta\mathcal{H}_+=\left(1+\frac{1}{4}\eta f\phi'^2|_{r_h}\right)T\delta\biggl(\frac{\mathcal{A}}{4}\biggr),
\end{equation}
where $\mathcal{A}$ is the area of the horizon. For this case, the differential $\left(1+\frac{1}{4}\eta f\phi'^2|_{r_h}\right)\delta\biggl(\frac{\mathcal{A}}{4}\biggr)$ cannot be interpreted as $\delta S$ for any function $S$ which is different from the neutral case \citep{Feng:2015oea}, because it is a variation in a two-dimensional parameter space. To resolve this issue, they take the two parameters to be the horizon radius $r_h$ and charge $Q$ of the black hole and introduce a new concept called `scalar charge' which is defined as
\begin{equation}
     Q_{\phi}=4\pi\sqrt{-\frac{8Q^2}{\eta r_{h}^2}}.
\end{equation}
With addition of $Q_{\phi}$, the $\delta \mathcal{H}_{+}$ becomes 
\begin{equation}
    \delta \mathcal{H}_{+}=T\delta S+\Phi_{\phi}\delta Q_{\phi},
\end{equation}
where $\Phi_{\phi}$ is the corresponding conjugate quantity of $Q_{\phi}$. 
Due to this additional term, the thermodynamic first law is different from the RN case which reads
\begin{equation}
    dM=TdS+\Phi_e dQ+\Phi_\phi dQ_\phi.
\end{equation}
The related thermodynamic quantities can be written as
\begin{equation}
\begin{aligned}
    &\Phi_{\phi}=-\frac18\eta r_{h}^2T\sqrt{-\frac{8Q^2}{\eta r_{h}^2}}, 
     Q_{\phi}=4\pi\sqrt{-\frac{8Q^2}{\eta r_{h}^2}},
    \\& \Phi_e=\frac{4 Q}{{r_h}}-\frac{8 Q^3}{3 r_h^3},\label{sclc}
\end{aligned}
\end{equation}
where $\Phi_e$ is the electric potential difference between the infinity and the black hole event horizon $\Phi_{e}=\Phi|_{\infty}-\Phi|_{r_{h}}$, which is calculated from the third equation of the Eq.(\ref{solution}) by choosing the electric potential at the infinity as zero. The condition $\eta<0 $ is required to ensure these thermodynamic variables are real.  Moreover, unlike Einstein gravity, the entropy of the black hole in Horndeski gravity has correction which is
\begin{equation}
    S=4\pi^2r_{h}^3T=\pi(r_{h}^2-2Q^2).
\end{equation}
From the expression of entropy, one can also find that $r_*<r_h$ guarantees the entropy $S$ to be positive.

While Ref.\cite{Feng:2015wvb} established the thermodynamic first law of charged Horndeski black hole by introducing the scalar charge $Q_{\phi}$, the physical meaning of additional scalar charge is still poorly understood. At least for asymptotic flat spacetime, a natural question is that if this scalar charge is an independent variable or not. This question is important since it will influence how the black hole evaporates. Thus in the next section, we will give an answer to this question and then investigate the evaporation behavior of Horndeski black hole.
\section{Evaporation behavior of charged black holes in Einstein-Horndeski gravity:}
As a black hole behaves like a thermal system, it can radiate
energy and evaporate. As an important part of black hole thermodynamics, we will discuss the evaporation behavior in this section.

In the thermodynamics of charged Horndeski black hole in Minkowski spacetime, besides the entropy and electric charge, there is an additional thermodynamic variable $Q_{\phi}$ which is called scalar charge. The evaporation will strongly depend on the properties of this scalar charge. Thus before investigating the evaporation behavior, it is important to illustrate the meaning of this scalar charge first.

In the case of $\alpha=0$ and $\Lambda=0$, the equation (\ref{fe}) reduces to 
\begin{equation}\label{con}
    \eta(rfh'+(f-1)h)=0.
\end{equation}
This condition is equivalent to $G^{rr}=0$, thus the rr component of Einstein
tensor should vanish for this black hole solution. It is interesting to see that the equation of motion of $\phi$ Eq.(\ref{eom1}) is trivially  satisfied
in this case because of $G^{rr}=0$ and $\alpha=0$
\begin{equation}
   (\alpha g^{rr}-\eta G^{rr})\phi' \equiv 0,
\end{equation}
which means the $\phi$ field can not be solved by its equation of motion. This is very different from other cases where scalar hair is involved such as Ref.\cite{Liu:2013gja,Lu:2014maa,Li:2020spf}. In these cases, the scalar field is solved by its equation of motion. It should be emphasized here that the Eq.(\ref{fe}) is important to escape from the constraints of no-hair theorem proposed in Ref.\citep{Hui:2012qt}. The proof of no hair theorem in Ref.\citep{Hui:2012qt} relies on the fact that the vanishing conserved current $J^{r}$ leads to vanishing scalar field $\phi$. However, by taking the condition Eq.(\ref{fe}) into account, the vanishing conserved current $J^r$ in Eq.(\ref{shl}) can not obtain the requirement of zero scalar field, which is crucial for the black hole to have non-trivial scalar hair as shown in Eq.(\ref{solution}). For more discussions on this point, the readers can consult Ref.\citep{Bravo-Gaete:2013dca,Babichev:2013cya,Jiang:2017imk,Zhang:2022hxl,Santos:2023flb,Santos:2023mee,Santos:2024cvx}. 



Moreover, for the Einstein-Horndeski action, it is interesting that the $\phi$ field only couples to curvature tensor by its derivative $\partial_\mu \phi$, which means the field has a shift symmetry
\begin{equation}
    \phi\rightarrow \phi +c.
\end{equation}
After gauging this symmetry to $\phi\rightarrow \phi+\lambda(x)$, Horndeski gravity can be seen as the Einstein-vector theory \citep{Geng:2015kvs} where $\partial_r \phi$ is treated as a vector field component $A_r$. In other words, the field $\phi$ can be eaten by the gauge field, thus the Einstein-Horndeski black hole solution in asymptotically Minkowski spacetime can be interpreted as the solution of non-minimally coupled Einstein-vector theory.

Let us stress more about its difference with respect to the Einstein-Maxwell theory. For a U(1) gauge field $A= \left(A_t(r), 0, 0, 0 \right)$, the Einstein-Maxwell theory predicts the RN black hole solution, with the electromagnetic field being static. However, it is interesting to ask how the metric changes if the electromagnetic field is slightly different, for example when an electromagnetic field takes the form $A = \left(A_t(r), A_r(r), 0, 0\right)$. If the gauge field is minimally coupled to the gravity, this has no effect on the black hole solution as this change will not affect the field strength $F_{\mu\nu}$ and stress-energy tensor. But if non-minimal coupling between vector field and metric is allowed, things will be different. For example, if the vector field component $A_{r}$ is non-minimally coupled to gravity in the Horndeski-like form
\begin{equation}\label{horn}
    L= R-\frac{1}{4}F^{2}+\frac{1}{2}\eta G^{rr}A_{r}A_{r},
\end{equation}
the metric and electromagnetic field outside the black hole will be drastically changed.\footnote{The vector field in Eq.(\ref{horn}) does not have $U(1)$ gauge symmetry, however, the gauge symmetry can emerge in the Horndeski black hole background (\ref{solution}) where $G^{rr}=0$ \cite{Geng:2015kvs}.} In this case, the equation of motion of $A_r$ field is $G^{rr}A_r=0$, if we impose constraint like Eq.(\ref{con}) where $G^{rr}=0$, there are no dynamics associated with $A_{r}$ which means $A_{r}$ can be chosen arbitrarily.
Asymptotically flat Horndeski black hole with constraints Eq.(\ref{con}) actually corresponds to the special choice of $A_r$ which is
\begin{equation}
    A_r=\phi^{\prime} =\sqrt{-\frac{8Q^2}{\eta r^2f}}.
\end{equation}
Therefore, the asymptotically flat Horndeski black hole solution in Eq.(\ref{solution}) is a good toy model to investigate the non-minimal coupling effect between the vector field and the metric field. From this point of view, we can conclude that the `scalar charge' introduced in the thermodynamic first law of Horndeski black hole \cite{Feng:2015wvb} is just the effect caused by vector field component $A_{r}$ which is not an independent variable in the thermodynamic system. It is determined by the electric charge $Q$ and mass $M$ with the relation Eq.(\ref{sclc}).\footnote{This needs to be distinguished from other cases involving scalar charges \cite{Lu:2014maa,Liu:2013gja,Li:2020spf}, where the scalar field has its own dynamics and associated scalar charges are the independent variables.} This interpretation also gives a physical reason for why the scalar charge is not included in neutral case \cite{Feng:2015oea}.  In the following, we will take this point of view of Horndeski black hole and further investigate its evaporation behavior. 

From the thermodynamic first law, the energy loss during the evaporation process consists of three parts. 
\begin{equation}
    \frac{dM}{dt}=T \frac{dS}{dt}+\Phi_{e}\frac{dQ}{dt}+\Phi_{\phi}\frac{dQ_{\phi}}{dt}.
\end{equation}
The first part is thermal Hawking radiation, which is the black-body radiation made up mostly of photons and neutrinos.
The second part is the charge loss of the black hole which should be governed by the Schwinger pair production effect caused by the electric field strength outside the horizon. The third part is the change of scalar charge, while as we discussed above, the scalar charge is not an independent variable that is solely determined by the mass and charge of the black hole.

The first term regarding thermal radiation follows Stefan-Boltzmann law 
\begin{equation}\label{sb}
    \frac{\delta Q_H}{dt}=T\frac{dS}{dt}=-\sigma T^{4}.
\end{equation}
Thus we need first to compute the cross-section $\sigma$. As the radiation is mostly made up of massless photons and neutrinos, the particles with long wavelengths are difficult to radiate to outside, so we take the high-frequency limit, and the radiation can be treated as null geodesics which is so-called the geometrical optics approximation \citep{Hiscock:1990ex, Page:1976df}. By these approximations, the cross-section for different particles becomes the geometric optics cross-section, thus the cross section in Eq.(\ref{sb}) is $\sigma=a b_{c}^{2}$ with $a$ the constant and $b_{c}$ the critical impact factor whose meaning will be explained later.

To compute what $b_{c}$ is, note that the trajectory of the mass-less particles obeys $ds^{2}=0$ which can be written more explicitly as  
\begin{equation}\label{geode}
    ds^{2}=-h \dot{t}^{2}+\frac{\dot{r}^{2}}{f}+r^{2}\dot{\phi}^{2}=0,
\end{equation}
here we take $\theta=\frac{\pi}{2}$ for simplicity and the dot is with respect to normalized affine parameter $\tilde{\lambda}$. Two conserved quantities can be defined along the null trajectory, 
\begin{equation}
    \mathcal{E}=h(r)\dot{t}, \quad L=r^{2}\dot{\phi},
\end{equation}
which is the energy and angular momentum of the massless particles respectively. 
By using these two conserved quantities and shifting the affine parameter to $\lambda=\tilde{\lambda} L$, the Eq.(\ref{geode}) can be rewritten as 
\begin{equation}
   \frac{dr}{d\lambda}=\sqrt{\frac{f(r)}{b^2 h(r)}-\frac{f(r)}{r^2}},
\end{equation}
where $b=L/\mathcal{E}$ is called the impact factor. A massless particle that can reach infinity must obey the equation 
\begin{equation}
    \frac{1}{b^2}\geqslant V(r),
\end{equation}
all along from horizon to infinity, where $V(r)=\frac{h(r)}{r^2}$ is regarded as effective potential. The critical impact factor can be obtained by the maximal value of effective potential
\begin{equation}
    b_c\equiv r_p/\sqrt{h(r_p)},
\end{equation}
where $r_p$ satisfies 
\begin{equation}
    \partial_{r}V(r)|_{r=r_{p}}=0,
\end{equation}
which is the radius of the unstable circular orbit of a massless particle. As shown in Ref.\citep{Gao:2023mjb}, it can be computed analytically as 
\begin{equation}
    \begin{aligned}
    r_{p}&=\frac{3 M}{4}+\frac{1}{2} [\frac{9 M^2}{4}+\frac{1}{3} \left( \root 3 \of {2} \delta +\frac{8\ 2^{2/3} Q^{4}}{\delta }-16 Q^2\right)]^{1/2}  \\&\quad +\frac{1}{2}[\frac{9 M^2}{2}-\frac{32 Q^2}{3}-\frac{\root 3 \of {2} \delta }{3}-\frac{8\ 2^{2/3} Q^4}{3 \delta }\\&+\frac{27 M^3-96 M Q^2}{4 \sqrt{\frac{9 M^2}{4}+\frac{1}{3} \left( \root 3 \of {2} \delta +\frac{8\ 2^{2/3} Q^4}{\delta }-16 Q^2\right) }}]^{1/2},\\
    \delta &=[Q^4 ( -243 M^2-3 \sqrt{6561 M^4-44928 M^2 Q^2+76800 Q^4}\\&+832 Q^2)]^{1/3}.
\end{aligned}
\end{equation}
So the equation of energy loss of Horndeski black hole reads
\begin{equation}
    \frac{\mathrm{d}M}{\mathrm{d}t}=-\frac{\pi^3}{15} b_c^2 T^4+\Phi_e \frac{\mathrm{d}Q}{\mathrm{d}t}+\Phi_{\phi} \frac{\mathrm{d}Q_{\phi}}{\mathrm{d}t},
\end{equation} 
where we choose $a$ to be the conventional value $\frac{\pi^{3}}{15}$. Note that we only care about its qualitative behaviors so the greybody factor is ignored.

\begin{figure*}[h]
    \centering
    \subfigure[Early stage of evaporation]{\includegraphics[width=0.5\textwidth]{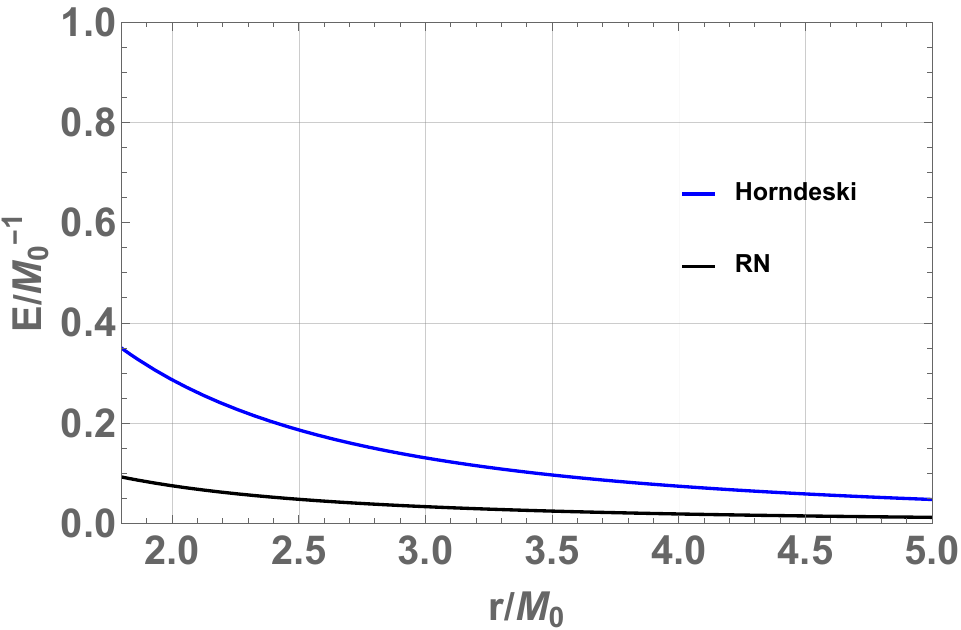}}\hfill
    \subfigure[Late stage of evaporation]{\includegraphics[width=0.5\textwidth]{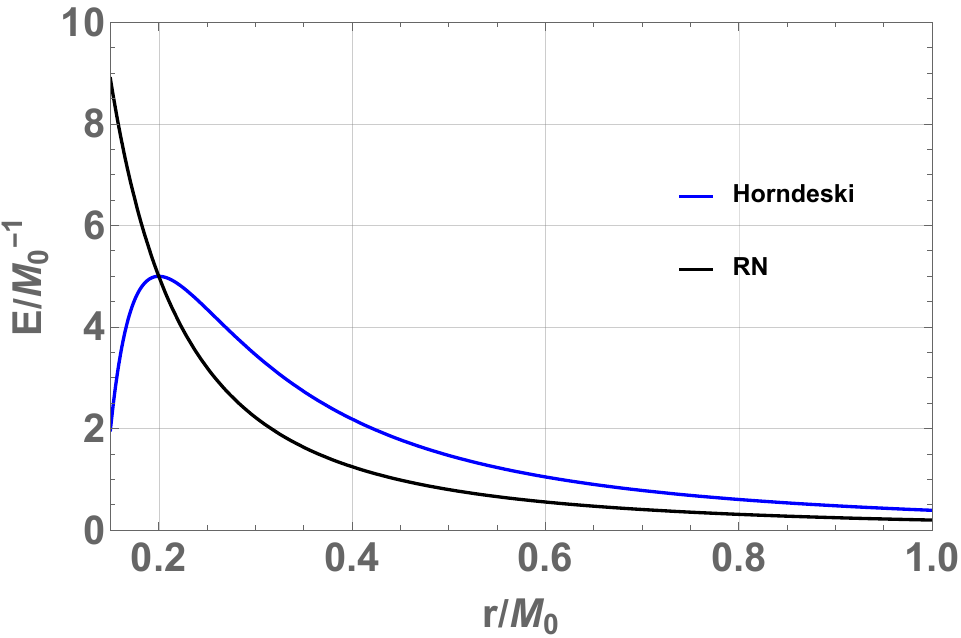}}
    \caption{Electric field strength outside the event horizon of RN and Horndeski black hole in different stages of evaporation. We choose the initial charge-mass ratio to be $\frac{Q_0}{M_0}=0.3$. The left panel represents the electric field strength at the early stage of evaporation (about $t/M_0\sim 10$) with the higher charge and larger horizon radius, while the right panel represents the case at the late stage of evaporation (about $t/M_0\sim 2100$) with the lower charge and smaller horizon radius. We find that the field strength outside the horizon between RN and Horndeski black hole will be qualitatively different in the late stage of evaporation.}
    \label{es}
\end{figure*}

Although scalar charge $Q_{\phi}$ is not an independent physical quantity as shown above, it obviously affects the evaporation process significantly. As $Q_{\phi}$ is an implicit function of $M$ and $Q$, we can rewrite the last term $\frac{\mathrm{d}Q_{\phi}}{\mathrm{d}t}$ as
\begin{equation}\label{qphi}
    \frac{\mathrm{d}Q_{\phi}}{\mathrm{d}t}=\frac{\partial Q_{\phi}}{\partial M} \frac{\mathrm{d}M}{\mathrm{d}t}+\frac{\partial Q_{\phi}}{\partial Q} \frac{\mathrm{d}Q}{\mathrm{d}t}.
\end{equation}
Then the energy loss equation can be rewritten as
\begin{equation}
     (1-\Phi_{\phi}\frac{\partial Q_{\phi}}{\partial M})\frac{\mathrm{d}M}{\mathrm{d}t}=-\frac{\pi^3}{15} b_c^2 T^4+(\Phi_e+ \Phi_{\phi}\frac{\partial Q_{\phi}}{\partial Q})\frac{\mathrm{d}Q}{\mathrm{d}t}.\label{dmdt}
\end{equation}
Then we only need to know the charge loss rate. The charge loss rate of the black hole is described by the Schwinger pair production effect \citep{Schwinger:1951nm}. We assume here that the radius of a black hole is much larger than the Compton wavelength of the electron, thus the production rate of electron-positron pairs is well described by the result of flat-space QED which can be written as \footnote{This formula is the same for many different kinds of matter field despite the difference between field strength \citep{Nayak:2005pf}. }\citep{Hiscock:1990ex,Xu:2019wak}
\begin{equation}
\Gamma=\frac{2^{}}{(2\pi)^3}m^4\left(\frac{E}{E_c}\right)^2\exp\left\{\frac{-\pi E_c}{E}\right\},\label{gamma}
\end{equation}
where $E_c\equiv \frac{m^2}{e}$ represent the Schwinger critical field strength below which the Schwinger effect is exponentially suppressed, and $m$, $e$ are electron mass and charge, respectively.\footnote{Although the vector field here is not necessarily electromagnetic field, the form of Eq.(\ref{gamma}) is the same for all kinds of vector fields such as dark matter field. As the lack of knowledge about the properties of dark matter, here we choose to use the mass and charge of electrons without loss of generality. The qualitative results will not change for other kinds of fields.} In Einstein-Horndeski theory, the electric field strength is given by
\begin{equation}
    E=-\frac{d\Phi}{dr}=\frac{4 Q}{r^2}-\frac{8 Q^3}{r^4}.\label{EC}
\end{equation}
For a more intuitive demonstration, we plot the electric field strength outside the horizon in Fig.\ref{es} for RN and Horndeski black holes respectively. In the left panel of Fig.\ref{es}, we show the field strength outside the horizon for the early stage of evaporation,  while in the right panel, we show field strength for the late stage of evaporation. To compare RN and Horndeski case better, we choose the same horizon radius $r_{h}$. As can be seen from Fig.\ref{es}, there is no qualitative difference in the field strength between the RN and Horndeski case in the early stage (both following $O(\frac{1}{r^{2}})$ Coulomb type ), but they are qualitatively different in the late stage. 

After the creation of charged particle-antiparticle pairs, one particle of the created pair can escape to spacetime infinity resulting in the loss of charge. Thus the charge loss rate of the black hole can be obtained by integrating the particle pair creation over the entire space outside the horizon, which is 
\begin{equation}
\begin{aligned}
     \frac{\mathrm{d}Q}{\mathrm{d}t}&=\int_{r_h}^{\infty}\int_{-\pi}^\pi\int_0^\pi \Gamma \sqrt{-g} dr d\theta d\phi\\&=-\int_{r_h}^{\infty}\frac{16 e^3 Q^2 \left(r^2-2 Q^2\right)^3 \exp\left\{\frac{\pi m^2 r^4}{8 e Q^3-4 e Q r^2}\right\}}{\pi ^2 r^8}\mathrm{d}r.\label{dqdt}
\end{aligned}
\end{equation}
The integral can only be solved numerically. Since we assume the black hole carries a positive charge $Q>0$, the emitted charged particles are actually positrons in our model. A similar discussion also holds for electrons, and so for simplicity, we shall use the term “electron” in the following discussions. 

\begin{figure*}[ht]
    \centering
    \subfigure[]{\includegraphics[width=0.33\textwidth]{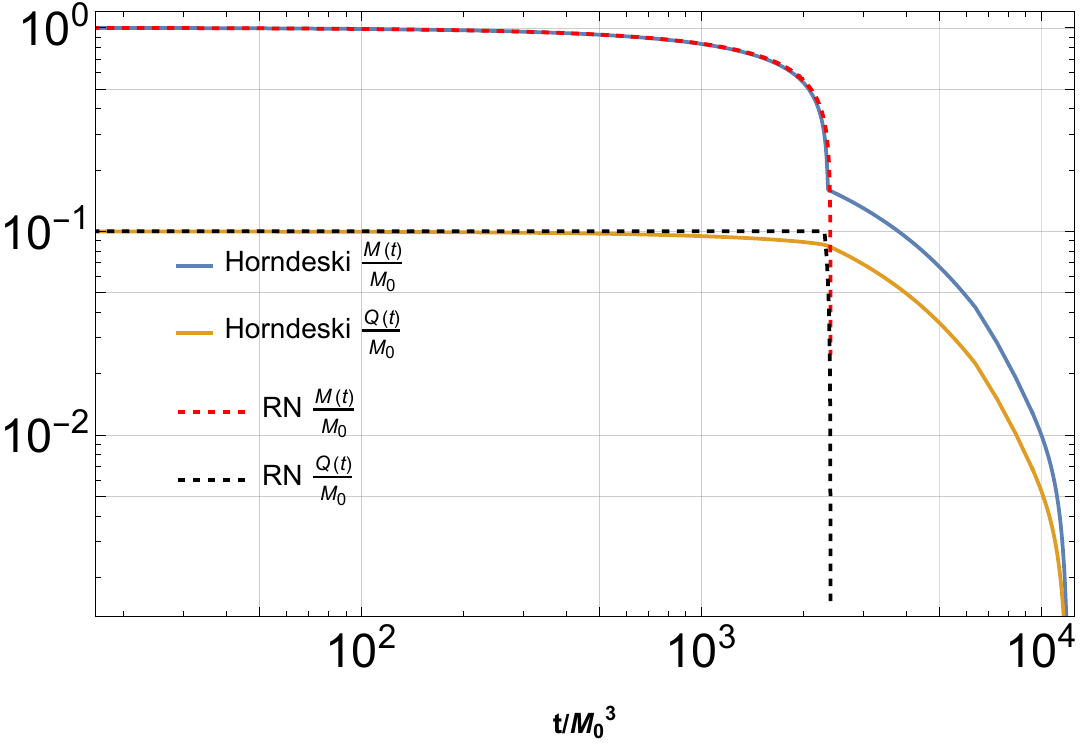}}\hfill
    \subfigure[]{\includegraphics[width=0.33\textwidth]{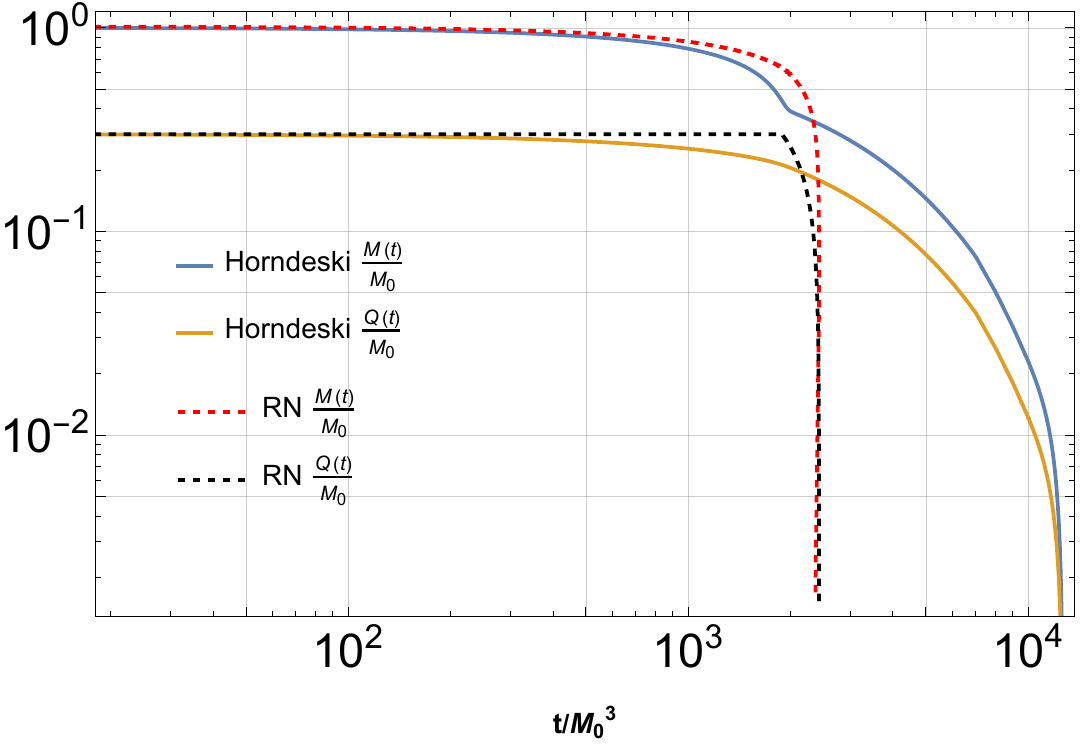}}
    \subfigure[]{\includegraphics[width=0.33\textwidth]{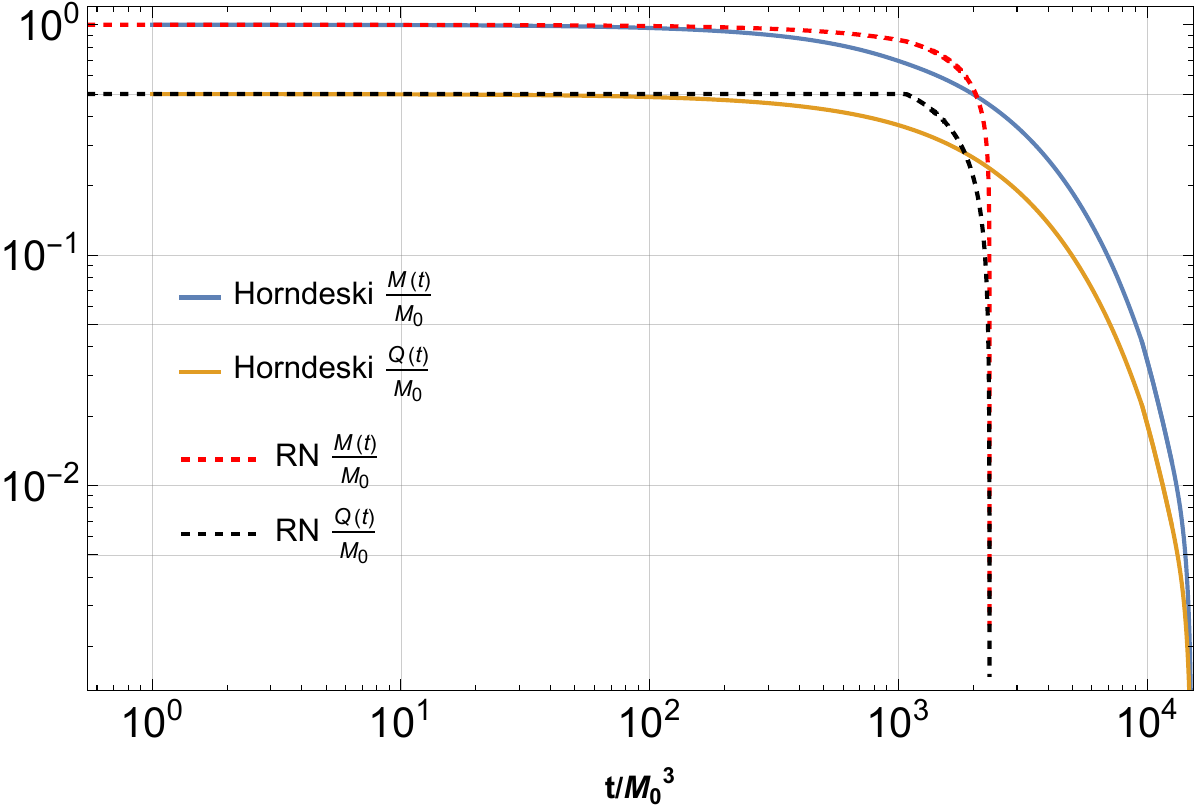}}
    \caption{The evolution of mass $M$ and charge $Q$ during evaporation process in charged Einstein-Horndeski black holes with different initial charge-mass ratios. The initial charge-mass ratios from left to right are $0.1 (a), 0.3 (b),$ and $0.5 (c)$ respectively. We set $\eta=-1$ without loss of generality. It is clearly seen that the evaporation behavior of the Horndeski black hole is different from the RN black hole in the late stage of evaporation which is after $t/M_{0}^{3}=$2400, 1988 and 1512 respectively.}
    \label{eva}
\end{figure*}

By solving the above system of ordinary differential equations (\ref{dmdt}) and (\ref{dqdt}), we numerically plot the evaporation process in Fig.\ref{eva} in which $M_0\sim10^7M_\odot$ with different initial charge-mass ratios. For an intuitive comparison with the RN case, we also plot the evaporation curves of RN black holes with the same initial mass and charge-mass ratio in the diagrams. The three diagrams show the evaporation process of three black holes with different initial charge-mass ratios, which are $Q_0/M_0=0.1, 0.3, 0.5$, respectively. As the black hole gradually loses its charge and mass during the evaporation process, the charge-to-mass ratio $Q/M$ gradually changes over time.  We can clearly see from the diagrams that in the case of Einstein-Horndeski charged black holes, in the early period the evaporation behavior of the first half is very similar to the RN case. However, when the horizon radius is decreased close to the position of singularity $r_{h}=\sqrt{2}Q$ which corresponds to the charge-mass ratio being $Q/M=\frac{3\sqrt{2}}{8}$, the evaporation rate takes a sudden turn and slows down due to the vanishingly small temperature here. After these points, the thermal radiation rate and charge loss rate become significantly smaller. This makes the lifetime of the Einstein-Horndeski charged black hole significantly longer than the RN black hole. Moreover, it needs to be stressed that during the evaporation, the charge-to-mass ratio always satisfies $Q/M<\frac{3\sqrt{2}}{8}$ which means the evaporation process can not violate cosmic censorship conjecture. 

To understand the difference in evaporation behavior between RN and Horndeski black hole, we see that in combination with (\ref{gamma}) and (\ref{dqdt}), the charge loss rate is closely related to the electric field strength outside the event horizon. At the early stage of evaporation, the horizon radius $r_h$ is large, thus $O(1/r^{4})$ term can be neglected in Eq.(\ref{EC}). Therefore, the electric field strength of Horndeski black holes outside the event horizon is the same as the RN case which is both $O(\frac{1}{r^{2}})$ Coulomb type. As the black hole evaporates, $r_h$ becomes smaller and the $O(\frac{1}{r^{4}})$ term of Horndeski electric field strength (\ref{EC}) becomes more and more important. As can be seen from Fig.\ref{es}, at the late stage of evaporation, the exterior electric field strength of the Horndeski black hole is vastly different from the RN case. The electric field strength outside Horndeski black hole first increases and then decreases instead of being a monotonically decreasing function. This difference in the electric field strength is the crucial reason for why the evaporation behavior of Horndeski black holes deviate from the RN black holes.

\section{Conclusion}

In this work, after reviewing the thermodynamics of the asymptotically flat charged Horndeski black hole and giving a physical understanding of the scalar charge $Q_{\phi}$, we investigate the evaporation behavior of this black hole. By solving the coupled differential equations of the mass loss and charge loss rate, we numerically calculate the evaporation process of this charged Einstein-Horndeski black hole. We demonstrated that the evaporation behaviors of Einstein-Horndeski black holes are very different from the RN case. 
The crucial reason for this difference is that the electric field outside the horizon is very different from the RN case at the late stage of evaporation, which makes the Schwinger pair production rate vastly different. This gives a Horndeski black hole a much longer evaporation lifetime than an RN black hole. The features of evaporation we found in this work may provide clues to distinguish Horndeski gravity from Einstein's gravity in future observations. 

For future directions, firstly, we only consider the asymptotically flat Horndeski black hole in this work. For Horndeski gravity, there are also black hole solutions in asymptotically AdS spacetime \cite{Feng:2015wvb}. Due to the presence of a time-like asymptotic boundary, Hawking radiation can be reflected back and equilibrium may be reached during the Hawking evaporation. There are some discussions on how the thermal equilibrium is reached for AdS black hole \cite{Page:2015rxa,Ling:2021nad}, it would be an interesting topic to discuss the effect of Horndeski coupling on this process. Moreover, for simplicity, we only consider the black hole evaporation in a special truncation of Horndeski gravity theory. The generalization of our analysis to more general scalar-tensor theories are still interesting to be pursued.  Finally, as we have shown in this paper, the scalar charge introduced in Horndeski gravity is not an independent variable because the scalar field does not have its own dynamics. This physical property is not very important in thermodynamical first law but can significantly affect the black hole evaporation behavior as evaporation is a dynamic process. Thus it would be interesting to investigate the evaporating behavior for other scalar hair black holes where the scalar hair has its own dynamics, such as the case Ref.\cite{Lu:2014maa,Liu:2013gja,Li:2020spf}. In these cases, the scalar charge will not follow Eq.(\ref{qphi}) during evaporation but has its own evolution equation based on the dynamics. This certainly will offer more interesting phenomena.   

\section*{Acknowledgements}
We are grateful to our group members for useful discussions. This work is supported by the National Natural Science Foundation of China (NSFC) under Grant Nos.12175105, 12405066, 11575083, and 11565017. Yu-Sen An is also supported by the Natural Science Foundation of Jiangsu Province under Grant No.BK20241376 and Fundamental Research Funds for the Central Universities. Ya-Peng Hu is also supported by the Top-notch Academic Programs Project of Jiangsu Higher Education Institutions (TAPP).


\bibliographystyle{elsarticle-num} 
\biboptions{sort&compress}
\bibliography{example}

\end{document}